\documentclass[conference]{IEEEtran}
\IEEEoverridecommandlockouts
\usepackage{cite}
\usepackage{amsmath,amssymb,amsfonts}
\usepackage{algorithmic}
\usepackage{graphicx}
\usepackage{textcomp}
\usepackage{xcolor}
\usepackage{booktabs}

\usepackage{subcaption} 

\def\BibTeX{{\rm B\kern-.05em{\sc i\kern-.025em b}\kern-.08em
		T\kern-.1667em\lower.7ex\hbox{E}\kern-.125emX}}
\usepackage{verbatim}
\usepackage[colorlinks,
linkcolor=blue,
anchorcolor=blue,
citecolor=blue]{hyperref}
\usepackage{amsmath,amssymb}
\usepackage{algorithmic,algorithm}

\usepackage{graphicx,textcomp,xcolor}
\usepackage{multirow, tabularx, ragged2e}
\newcolumntype{C}{>{\Centering\arraybackslash}X}

\usepackage{float}

\begin{document}

\title{Enhancing Portfolio Optimization with Transformer-GAN Integration: A Novel Approach in the Black-Litterman Framework}

\author{Enmin Zhu, Jerome Yen}

\maketitle

\begin{abstract}
	This study presents an innovative approach to portfolio optimization by integrating Transformer models with Generative Adversarial Networks (GANs) within the Black-Litterman (BL) framework. Capitalizing on Transformers' ability to discern long-range dependencies and GANs' proficiency in generating accurate predictive models, our method enhances the generation of refined predictive views for BL portfolio allocations. This fusion of our model with BL's structured method for merging objective views with market equilibrium offers a potent tool for modern portfolio management, outperforming traditional forecasting methods. Our integrated approach not only demonstrates the potential to improve investment decision-making but also contributes a new approach to capture the complexities of financial markets for robust portfolio optimization.
\end{abstract}

\begin{IEEEkeywords}
Transformer, GAN, Black and Litterman model, Portfolio optimization
\end{IEEEkeywords}

\section{Introduction}

The Black-Litterman (BL) model, foundational in modern portfolio theory, adeptly merges market equilibrium with investor insights to refine portfolio allocations. This integration addresses key limitations of the traditional Markowitz mean-variance optimization, particularly issues related to input sensitivity and estimation errors\cite{black1990asset}\cite{doi:10.2469/faj.v48.n5.28}. The generation of accurate and objective investor views remains a significant challenge, underlining the necessity for advanced, more effective methods.

Recent advancements in artificial intelligence provide promising solutions for generating these views. CNNs and LSTMs have been instrumental within the BL framework, leveraging a range of data indicators and historical price analyses to enhance portfolio performance\cite{selvin2017stock}\cite{rezaei2021intelligent}\cite{barua2022dynamic}. However, their effectiveness is often hampered by the complexity of processing multiple indicators and their struggles with long-term dependencies\cite{radford2018improving}. These challenges highlight the need for innovative and robust approaches that can better capture and utilize the intricacies of financial data.

Enter the Transformer and Generative Adversarial Network (GAN) models, which represent the cutting edge of deep learning technology. Originally developed for natural language processing, Transformers have shown remarkable versatility and superior performance in various domains, including computer vision\cite{dosovitskiy2020image} and speech recognition\cite{fan2021improved}, due to their unique architecture that facilitates effective sequential data handling\cite{radford2018improving}. Similarly, GANs have emerged as powerful tools for modeling and generating complex data distributions, proving particularly effective in unveiling latent patterns within historical financial data\cite{NIPS2014_5ca3e9b1}.

This study introduces a novel integration of GANs with Transformer models within the Black-Litterman framework to optimize portfolio construction. This hybrid approach not only enhances the generation of investor views by distilling complex financial time series data into actionable insights but also significantly improves the predictive accuracy and robustness of portfolio allocations. Our contributions can be summarized in three main advancements:

\begin{itemize}
    \item \textbf{Advanced Architectural Integration:} Our model uniquely combines the generative capabilities of GANs with the deep learning efficiency of Transformers, making it particularly adept at uncovering long-range dependencies and latent patterns within historical stock price data.
    \item \textbf{Performance and Adaptability:} Through rigorous benchmarking, our model demonstrates superior accuracy and scalability, adapting seamlessly to various data resolutions and market conditions.
    \item \textbf{Strategic Asset Allocation:} The integration of our model with the Black-Litterman framework allows for a more nuanced and adaptive approach to portfolio construction, significantly enhancing the potential to outperform traditional market strategies.
\end{itemize}

\section{Related Work}

\subsection{Enhancements in Portfolio Optimization Models}

The Mean-Variance (MV) model, foundational in portfolio optimization, has notable limitations, particularly its sensitivity to input estimates which can lead to non-robust portfolio allocations \cite{bessler2017multi}. These portfolios often lack true diversification and may concentrate heavily in a few assets. Moreover, the MV model struggles with the asymmetry of stock returns and often underestimates risk during market downturns. To address these issues, more robust models such as the Black-Litterman (BL) model have been developed. The BL model integrates market equilibrium with investor views to enhance portfolio construction \cite{teplova2023black}, aiming for more intuitive and stable asset allocation \cite{adelmann2016improvement}. This model addresses the limitations of the MV model by generating portfolios with lower risk, less extreme asset allocations, and higher diversification across asset classes \cite{fabozzi2006incorporating}.

\subsection{Deep Learning Approaches in Financial Markets}

Deep learning algorithms, particularly Convolutional Neural Networks (CNNs), utilize convolutional kernels to autonomously extract pivotal features from stock price charts or technical indicators, achieving remarkable accuracy \cite{sezer2020financial} \cite{wu2020cnn}. However, CNNs do not inherently account for the sequential nature of stock prices, necessitating additional transformations of time-series data \cite{vidal2020gold}. Addressing this, Long Short-Term Memory networks (LSTM), a specialized subset of Recurrent Neural Networks (RNN), employ mechanisms to modulate information flow, significantly boosting their efficacy in processing time-series data in stock analysis \cite{sezer2020financial} \cite{di2017recurrent}. Despite their prowess in temporal feature capture, LSTMs are less adept at extracting features from high-dimensional spatial data. The CNN-LSTM hybrid model has been proposed to merge CNNs' robust feature extraction capabilities with LSTMs' proficiency in time-series analysis \cite{vidal2020gold} \cite{lu2020cnn} \cite{lu2021cnn}. Further advancements include integrating the attention mechanism into the CNN-LSTM framework, enhancing prediction accuracy and efficiency \cite{wu2021attention} \cite{sun2021two} \cite{zhang2023stock}.

\subsection{Transformer and Attention Mechanisms}

The attention mechanism, pivotal to the Transformer model, has revolutionized the field of sequential data analysis by enhancing the model's focus on relevant segments of input data, particularly beneficial for processing lengthy sequences \cite{vaswani2017attention}. This innovation allows Transformers to process data in parallel, handle long-range dependencies, and scale efficiently, enhancing prediction accuracy and the modeling of complex patterns. Its effectiveness is demonstrated in stock market forecasting, where its self-attention mechanism captures long-term dependencies \cite{wang2022stock} \cite{zhang2022transformer}. In comparison, LSTM models, due to their limited ability in capturing long-distance dependencies, often fall short in performance metrics such as accuracy and training efficiency \cite{radford2018improving}. This limitation is also noted in CNN models, which struggle with processing long-term temporal dependencies \cite{zeng2023financial}. The Transformers' self-attention mechanism effectively weighs the importance of all parts of the input data, irrespective of their position \cite{manning2020emergent} \cite{ameri2023impact} \cite{ferrando2023explaining}. Sun et al. employ a modified Transformer network, stripped of the position encoding module, to learn the policy for determining the subjective views of expected return for the Black-Litterman model \cite{sun2024combining}.

\subsection{Advancements in Generative Models for Financial Analysis}

Generative Adversarial Networks (GANs) stand out for their unique operational framework, predicated on the concept of a two-player zero-sum game \cite{NIPS2014_5ca3e9b1}. These models have shown remarkable capability in capturing potential patterns or features from complex datasets \cite{hong2019generative}. For instance, GANs have been used in anomaly detection, showcasing their capacity to identify abnormal instances by learning potential patterns within data \cite{rani2020survey}. GANs have also shown promise in the domain of time series forecasting, with models like StockGAN demonstrating efficacy in financial time series analysis \cite{diqi2022stockgan}.

\section{Background}
In the realm of machine learning, various architectures have been developed to tackle a wide range of tasks, each with unique strengths and applications. This section provides an overview of four pivotal models: Convolutional Neural Networks (CNNs), Long Short-Term Memory networks (LSTMs), Generative Adversarial Networks (GANs), and Transformers. CNNs are renowned for their image processing prowess, LSTMs for their ability to capture time-dependent information, GANs for their generative capabilities in producing new, synthetic instances of data, and Transformers for their advanced handling of sequential data through attention mechanisms. We will discuss the key features, foundational principles, and the specific roles these models play in the context of our research.

\subsection{GAN}
Generative Adversarial Networks (GANs) feature two neural models—the Generator (G) and the Discriminator (D)—in a competitive setup that produces high-quality synthetic data closely mimicking real distributions. This section explains their roles and the mathematics underlying their interaction.

\begin{figure}[H]
	\centering
	\includegraphics[scale=0.3]{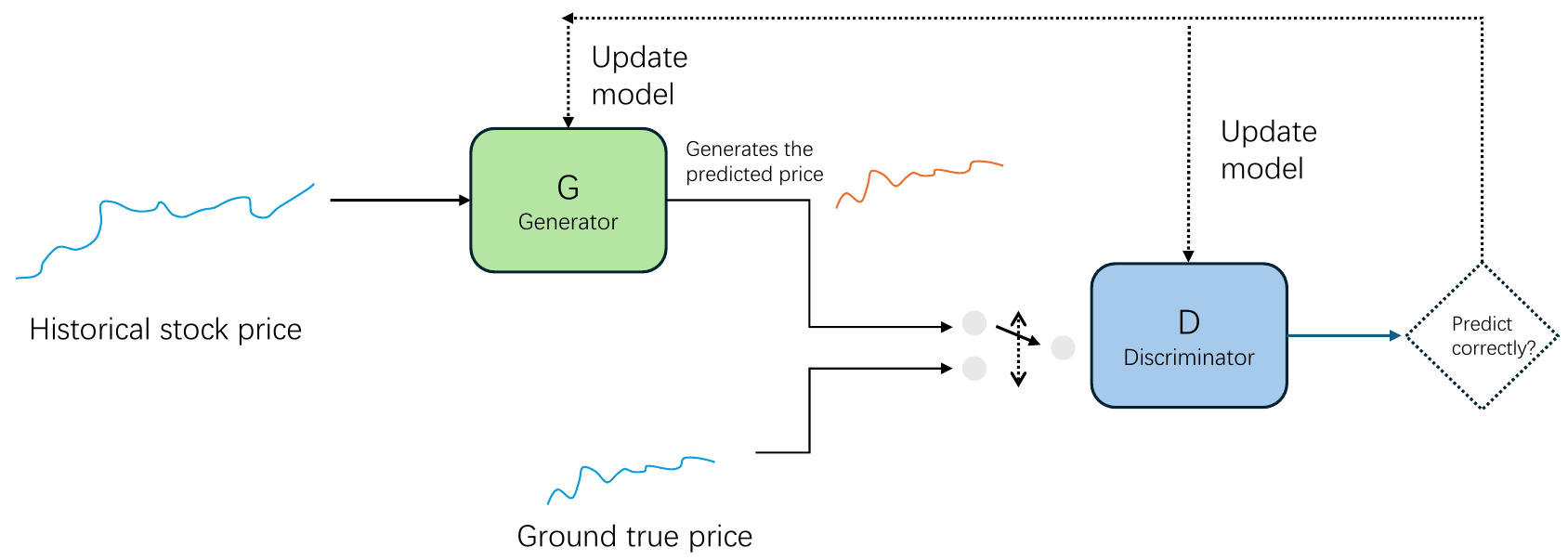}
	\caption{Structure of Generative Adversarial Network(GAN).}
	\label{gan}
\end{figure}

A GAN consists of two main components: the Generator (G) and the Discriminator (D), which are refined through adversarial learning. The Generator's role is to create new, unique input samples that mimic the structure of real data, starting from random noise, despite having no prior access to the real data itself. Conversely, the Discriminator evaluates these samples, distinguishing between the authentic and generated data, aiming to identify disparities linked to the origin of the data. Fig. \ref{gan} visualizes this process, with 'Z' indicating the input noise vector, 'G(Z)' the data generated by the Generator, 'X' the real data, 'G' the Generator, and 'D' the Discriminator.
In GAN, both models, denoted as $(G, D)$, are engaged in a continual process of mutual learning, a topic we will delve into in the next section. The GAN model is represented by individual neural networks, with each network operating in opposition to the other. Specifically, $x$ is a sample drawn from the real data distribution $P_{data}(x)$, while $z$ is sampled from a prior distribution $p_z(z)$, such as a uniform or Gaussian distribution. The symbol E(·) represents the expectation operator. $D(x)$ represents the probability of $x$ being sampled from the real data rather than being generated. When the input data is from real data, the discriminator aims to maximize $D(x)$, approaching the value of 1. Conversely, when the input data is generated by the generator $G(z)$, the discriminator endeavors to make $D(G(z))$ approach 0 and generator $G$ tries to make it to 1.

Given the competitive nature of this zero-sum game involving two players, the optimization problem of GAN can be formulated as a minimax problem or the lossing function:

\begin{equation}
	\resizebox{0.9\hsize}{!}{$
		\begin{aligned}
			\min_{\substack{G}} \max_{\substack{D}} V(D,G) =  \mathbb{E}_{\boldsymbol{x}\sim p_{\text{data}}(\boldsymbol{x})}[\log D(\boldsymbol{x})] + 
			\mathbb{E}_{\boldsymbol{z}\sim p_{\boldsymbol{z}}(\boldsymbol{z})}[\log(1-D(G(\boldsymbol{z})))]
		\end{aligned}
		$}
\end{equation}

In summary, when training Generative Adversarial Networks (GANs), our objective is to learn the model's parameters through competitive training of the generator and discriminator. The generator's task is to produce realistic data, making it difficult for the discriminator to distinguish, while the discriminator aims to accurately differentiate between real and generated data. During the training process, the generator attempts to deceive the discriminator, while the discriminator strives to enhance its ability to discriminate between real and generated data. Through this competition, the generator and discriminator of GANs gradually improve their performance, ultimately achieving the goal of generating high-quality and realistic data.

\subsection{Transformer Mechanism}
The Transformer architecture, introduced by Vaswani et al.\cite{vaswani2017attention}, has revolutionized the field of deep learning by offering an alternative to recurrent neural networks (RNNs) for processing sequential data. Central to the Transformer's architecture is the self-attention mechanism, which enables the model to weigh the significance of different parts of the input data without the sequential processing limitations of RNNs.

The key innovation of the Transformer is its ability to handle dependencies between any two elements in the input data, regardless of their positions. This capability is achieved through the self-attention mechanism, which computes a weighted sum of all input elements, with weights assigned based on a function of the elements' compatibility. The mathematical representation of the self-attention mechanism is given by:
\begin{equation}
    \text{Attention}(Q, K, V) = \text{softmax}\left(\frac{QK^T}{\sqrt{d_k}}\right)V
\end{equation}
where $Q$, $K$, and $V$ represent the queries, keys, and values matrices respectively, and $d_k$ is the dimensionality of the keys.

Each element in the output sequence of a Transformer layer is computed as a weighted sum of all values, with the weights assigned according to the dot products of the query with all keys. This allows each output element to dynamically attend to all input elements, facilitating better learning of contextual relationships in the data.

The Transformer uses multi-head attention to extend this mechanism:
\begin{equation}
    \text{MultiHead}(Q, K, V) = \text{Concat}(\text{head}_1, \dots, \text{head}_h)W^O
\end{equation}
\begin{equation}
    \text{where head}_i = \text{Attention}(QW_i^Q, KW_i^K, VW_i^V)
\end{equation}

Here, $W^O, W_i^Q, W_i^K, \text{and } W_i^V$ are parameter matrices, and $h$ is the number of attention heads. This design allows the model to jointly attend to information from different representation subspaces at different positions, enhancing the model's ability to learn complex patterns.

Transformers are structured into an encoder-decoder architecture. The encoder maps an input sequence to a continuous representation that holds all the learned information of that sequence. The decoder then generates an output sequence from this representation. The entire process is facilitated by residual connections and layer normalization, enhancing training stability and model performance.

\subsection{Black-Litterman Framework}
The Black-Litterman (BL) model refines traditional portfolio optimization by blending market data with investor views. This integration improves investment strategies by aligning market conditions with personal risk preferences. Below are the key formulas that illustrate this method's approach to asset allocation.

Formula (\ref{sig}) represents the calculation of the equilibrium return vector, $\Pi$, where $\delta$ denotes the risk aversion coefficient, $\Sigma$ is the covariance matrix of asset returns, and $w_{marker}$ are the market capital weights. 
\begin{equation}
        \label{sig}
	\Pi=\delta\Sigma w_{market}
\end{equation}

$\Omega$ quantifies the uncertainty associated with the investor's views. It is a diagonal matrix where each diagonal element represents the confidence or variance around each view expressed by matrix $P$. The term $\tau$ is a scalar that reflects the uncertainty of the equilibrium return vector, $\Sigma$ is the covariance matrix of returns, and $P$ is a matrix that maps each view onto the assets.
\begin{equation}
	\label{omega}
	\Omega=\operatorname{diag}(\mathsf{P}(\tau\Sigma)P^T)
\end{equation}

This equation adjusts the original covariance matrix, $\Sigma$, to account for the additional uncertainty introduced by the investor's views. The adjusted covariance matrix, $\hat{\Sigma}$, incorporates both the market's inherent volatility and the specific risk perceptions associated with the investor's views.
\begin{equation}
	\hat{\Sigma}=\Sigma+\left(\left(\tau\Sigma\right)^{-1}+\mathrm{P^T\Omega^{-1}P}\right)^{-1}
\end{equation}

$\mu_{BL}$ denotes the Black-Litterman expected returns, which combine the market equilibrium returns with the investor's views. This is achieved by adjusting the market equilibrium returns, $\Pi$, with the investor's specific views on returns, $Q$, taking into account the confidence in these views as captured by $\Omega$.
\begin{equation}
	\label{bl}
	\mu_{BL}=\left[\left(\tau\Sigma\right)^{-1}+P^{'}\Omega^{-1}P\right]\left[\left(\tau\Sigma\right)^{-1}\Pi+P^{'}\Omega^{-1}Q\right]
\end{equation}

\begin{equation}
	\label{weight}
	\max_{w} \frac{(w^T \mu_{BL}) - r_f}{\sqrt{w^T \hat{\Sigma} w}} 
\end{equation}
\begin{equation}
	\label{w>0}
	\text{subject to: } \sum_{i=1}^{n} w_i = 1, \quad w_i \geq 0
\end{equation}

As is shownd in Formulas (\ref{weight}) and (\ref{w>0}), the maximization of the Sharpe ratio is formulated as a convex optimization problem to determine the optimal asset weights in a portfolio. This formulation adapts to the Black-Litterman model's outputs, taking into account the expected returns and the adjusted covariance matrix. In line with common institutional investment practices, our portfolio construction excludes negative weighting to reflect the general avoidance of short selling by institutional investors\cite{barua2022dynamic}. Hence, the Sharpe ratio maximization is subject to the constraints that the sum of the weights equals one and that all weights are non-negative. This technique ensures that the portfolio construction is not only aligned with the investor's views but also optimized for risk-adjusted returns.

In evaluating the effectiveness of these methods, it is essential to consider factors such as accuracy in predicting market trends, adaptability to market changes, and the ability to mitigate biases inherent in manual view generation. Studies have shown that portfolios constructed using deep learning-enhanced Black-Litterman models tend to outperform those based on traditional weight allocation schemes. This suggests a significant potential for advanced methods, especially deep learning algorithms, to improve the efficacy of the Black-Litterman model in contemporary portfolio management.

\section{Methodology}
In this section, we delve into the BL-TGAN (Black-Litterman Transformer-GAN) architecture, a sophisticated architecture that marries the predictive prowess of Transformer networks with the generative capabilities of Generative Adversarial Networks (GANs), all within the Black-Litterman framework. The Transformer’s role is to analyze time-series data for feature extraction and trend prediction, while the GAN is tailored to refine these predictions, ensuring that the generated investment views are not just a reflection of past patterns but also adaptable to new market dynamics.

The methodology's phases are concisely depicted in Fig. \ref{structure} below.

\begin{figure*}[htbp]
	\centering
	\includegraphics[scale=0.7]{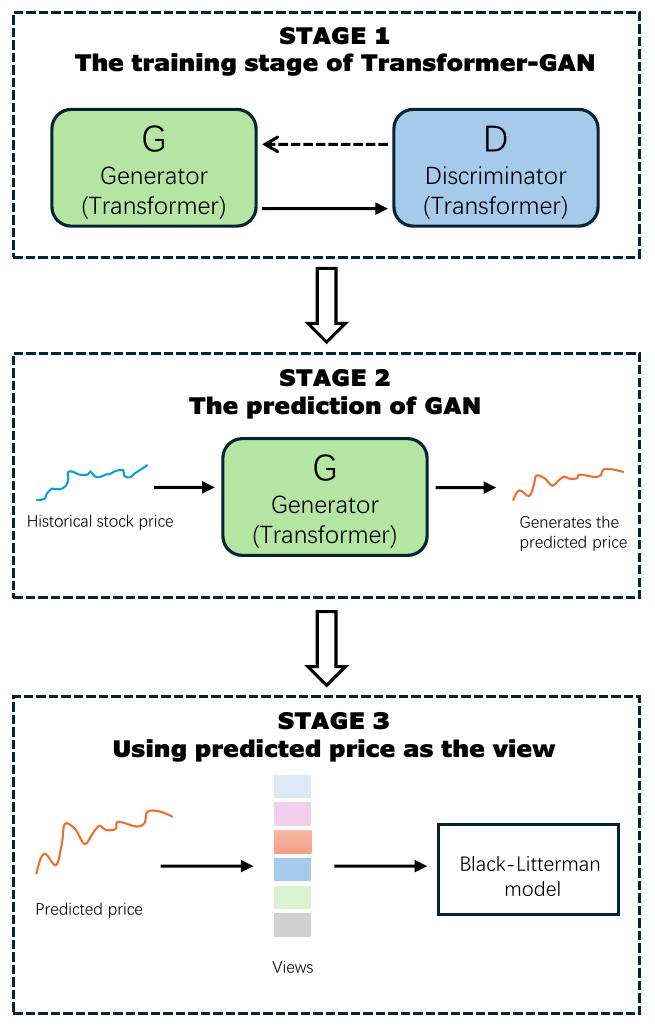}
	\caption{Architecture of our portfolio construction method}
	\label{structure}
\end{figure*}

\subsection{Hybrid Combination of Transformer and GAN}
Transformer-GAN is the combination of transformer model and GAN architecture. The innovative approach of combining Transformer models and Generative Adversarial Networks (GANs) in the context of financial time series prediction, specifically for predicting stock closing prices, leverages the mechanism of GANs—setting generator and discriminator as Transformer models—rather than generating synthetic data. This novel method enhances prediction accuracy and reliability by integrating the strengths of both architectures in a unique manner.

\subsubsection{Adaptation of GAN Mechanism}
Contrary to conventional uses of GANs, which are typically aimed at creating synthetic data that mimics real datasets, our research repurposes the GAN architecture to directly forecast stock closing prices. 

In our novel approach, the GAN generator is designed to generate predictions of stock prices, while the discriminator evaluates the accuracy of these predictions against actual historical data. This paradigm shift leverages the adversarial process of GANs to fine-tune the generator's ability to forecast prices, with the discriminator acting as a dynamic benchmark that guides the generator towards more precise estimations. This adversarial training approach allows the generator to learn and improve its predictions over time, resulting in more accurate and realistic future stock price predictions\cite{diqi2022stockgan}\cite{ma2024vgc}.

\subsubsection{The Superior Features of Transformer Models}
The transformer model, a revolutionary architecture in machine learning, operates on principles that enable it to process sequential data efficiently. Central to its functionality are the concepts of attention mechanisms\cite{manning2020emergent}\cite{ameri2023impact}\cite{ferrando2023explaining} and positional encoding\cite{shiv2019novel}\cite{kazemnejad2024impact}, which allow it to handle dependencies and understand the order in sequential data like financial time series.

Positional encoding is added to the input embeddings to give the model information about the position of the words in the sequence. The positional encoding can be represented as:
\begin{equation}
	PE_{(pos, 2i)} = \sin\left(\frac{pos}{10000^{2i/d_{\text{model}}}}\right)
\end{equation}
\begin{equation}
	PE_{(pos, 2i+1)} = \cos\left(\frac{pos}{10000^{2i/d_{\text{model}}}}\right)
\end{equation}
In these equations, $pos$ is the position in the sequence, and $i$ is the dimension.

The utilization of transformer models in this study is pivotal for generating informed and data-driven views within the Black-Litterman framework. The implementation process is outlined in distinct phases, ensuring a comprehensive and meticulous approach.

\subsubsection{Transformer-GAN Working Mechanism}
In this hybrid model, while the transformer serving as the generator focuses on analyzing historical stock prices to predict future closing prices, the discriminator transformer model evaluates the accuracy of these predictions against real historical data. This competitive setup encourages the predictive model to generate increasingly accurate forecasts, thereby improving its ability to capture the underlying trends and patterns in the data.

The benefits of this hybrid approach include:
\begin{itemize}
	\item\textbf{Enhanced Predictive Accuracy}: The iterative refinement process between the generator and discriminator transformers leads to improved prediction accuracy. The discriminator's challenge compels the generator to produce outputs that are closely aligned with real market behaviors.
	\item\textbf{Robustness to Market Volatility}: The model's exposure to continuous evaluation and challenge enhances its robustness, making it better equipped to handle market volatility and adapt to emerging patterns not explicitly present in the training data.
	\item\textbf{Capturing Latent Market Signals}: GANs are exceptionally adept at distilling latent information within stock price data, information that often eludes traditional analytical models. The generator component of a GAN learns, through its training process, to predict future stock prices by reflecting complex market signals. This enables the effective unveiling of patterns predictive of future market movements.
\end{itemize}


\subsection{Generating Views with Transformer-GAN architecture}

In our approach, the Transformer-GAN architecture processes historical data to forecast future asset returns. The predicted prices $p_t$ are generated from historical prices $p_{t-1}, \ldots, p_{t-5}$. The weekly return will be calculated by the predicted price and structured into a view vector $Q$, with each element outlining a return prediction for an asset. The views are linked to their respective assets via the construction of matrix $P$, while the confidence in each view is quantified by the diagonal matrix $\Omega$, typically based on historical volatility.

These elements are integrated into the Black-Litterman model, which are shown in Table \ref{tab:BL_symbols}, adjusting market equilibrium returns with the Formula (\ref{bl}). Here, $\mu_{BL}$ denotes the adjusted returns, $\Pi$ the market equilibrium returns, $P$ the matrix mapping views onto assets, $Q$ the vector of views, $\Sigma$ the covariance matrix of returns, $\Omega$ the confidence in the views, and $\tau$ the scalar adjusting market covariance. In our method, $\tau$ is 0.025. The resulting posterior returns $\Pi$ are then utilized in a mean-variance optimization to compute the optimal portfolio weights, balancing the Sharpe ratio against risk.

\begin{table}[htbp]
	\centering
	\caption{Mathematical Symbols in the Black-Litterman Model}
	\label{tab:BL_symbols}
	\begin{tabular}{@{}cl@{}}
		\toprule
		Symbol & Meaning \\
		\midrule
		\( \mu_{BL} \) & Adjusted returns \\
		\( \Pi \) & Market equilibrium returns \\
		\( P \) & Matrix mapping views onto assets \\
		\( Q \) & Vector of views \\
		\( \Sigma \) & Covariance matrix of returns \\
		\( \Omega \) & Confidence in the views \\
		\( \tau \) & Scalar adjusting market covariance (which is 0.025 here) \\
		\bottomrule
	\end{tabular}
\end{table}

\subsection{Assessment Metrics}

In evaluating the performance of predictive models, particularly those applied in financial markets like the transformer model for generating views in the Black-Litterman framework, it is crucial to employ accurate and insightful assessment metrics. This study utilizes a suite of metrics to quantify the predictive accuracy and reliability of the model outputs, including Mean Squared Error (MSE), Mean Absolute Error (MAE), and Normalized Mean Squared Error (NMSE). These metrics offer a comprehensive view of model performance, each highlighting different aspects of prediction errors.

\begin{itemize}[]
\item\textbf{Mean Squared Error (MSE)}\\
The Mean Squared Error (MSE) is a standard metric used to measure the average of the squares of the errors, that is, the average squared difference between the estimated values ($\widehat{y}_i$) and the actual values ($y_i$). It is defined mathematically as:
\begin{equation}
	\text{MSE} = \frac{1}{\zeta}\sum_{i=1}^{\zeta}\left(y_i-\widehat{y}_i\right)^2
\end{equation}
where $\zeta$ represents the total number of observations. MSE is particularly useful for quantifying the magnitude of the prediction error, with lower values indicating better model performance. However, it is sensitive to outliers as it squares the errors before averaging.

\item\textbf{Mean Absolute Error (MAE)}\\
The Mean Absolute Error (MAE) measures the average magnitude of the errors in a set of predictions, without considering their direction. It is calculated as the average of the absolute differences between predicted and actual values:
\begin{equation}
	\text{MAE} = \frac{1}{\zeta}\sum_{i=1}^{\zeta}\left|y_i-\widehat{y}_i\right|
\end{equation}
MAE provides a straightforward interpretation of prediction errors and, unlike MSE, is not unduly influenced by outliers. Lower MAE values signify more accurate predictions.

\item\textbf{Normalized Mean Squared Error (NMSE)}\\
The Normalized Mean Squared Error (NMSE) offers a way to compare the model's error relative to the variance of the dataset, thereby providing a normalized measure of prediction accuracy:
\begin{equation}
	\text{NMSE} = \frac{\sum_{i=1}^{\zeta}\left(y_i-\widehat{y}_i\right)^2}{\zeta^*\text{var}(y_i)}
\end{equation}
In this equation, $\text{var}(y_i)$ denotes the variance of the actual values. NMSE adjusts the MSE by the variance of the data, making it a relative measure of model performance. Values closer to zero indicate better model accuracy, with negative values suggesting performance worse than the simple mean of the target variable.
\end{itemize}

\section{Experiment}

This section presents a detailed experimental study, our study designs a series of experiments employing real-world financial data. We outline the data sources, the preprocessing steps taken to ensure data quality, and the rigorous evaluation criteria we will apply to assess the model's performance. Through these experiments, we anticipate showcasing our model’s ability to enhance portfolio construction and optimization based on black-litterman framework, thus offering valuable contributions to the domain of quantitative finance.

\subsection{Experiment Setup}
The experiment comprises several steps, ranging from data simulation to applying the Transformer-GAN model, followed by integrating its output into the Black-Litterman model, and concluding with portfolio optimization. 

\subsection{Data Collection and Processing}

\subsubsection{Data Source}
As discussed, the dataset provided by Kaggle\cite{equinxx2023} encompasses historical prices of assets, trading volumes, and various financial metrics. This dataset also includes tweets for the top 25 most-watched stock tickers on Yahoo Finance, spanning from 30-09-2021 to 30-09-2022. This dataset serves as a critical resource for training the transformer models, as well as for other models employed in our controlled experiments. These experiments aim to test the effectiveness of the models in a practical setting.

\subsubsection{Data Preprocessing}
We normalize the stock prices to a range of (0,1) and revert them to their original scale upon obtaining results, which are then utilized to generate views for the Black-Litterman model. The data is partitioned into training and testing sets with proportions of 0.8 and 0.2, respectively. These preprocessing steps are crucial to ensuring the integrity and reliability of our model training and analysis. Subsequently, we conducted a statistical analysis of samples from dataset, the results of several tickers which are presented in Table \ref{tab:stocks_samples}.

\begin{table}[H]
	\centering
	\caption{Statistic Information of Sample from Our Stock Pool(2021-2022)}
	\label{tab:stocks_samples}
	\begin{tabular}{lccccc}
		\hline
		\textbf{Stock} & \textbf{Count} & \textbf{Mean} & \textbf{Max} & \textbf{Min} & \textbf{Std} \\ \hline
		ENPH           & 251            & 0.003562      & 0.246512     & -0.123061    & 0.047608     \\
		BX             & 251            & -0.000793     & 0.084490     & -0.100033    & 0.031066     \\
		F              & 251            & -0.000347     & 0.116674     & -0.123242    & 0.031439     \\
		MSFT           & 251            & -0.000491     & 0.066852     & -0.054978    & 0.019609     \\
		GOOG           & 251            & -0.000985     & 0.077390     & -0.058640    & 0.021730     \\
		AAPL           & 251            & 0.000228      & 0.069778     & -0.058680    & 0.020061     \\
		VZ             & 251            & -0.001248     & 0.043521     & -0.067352    & 0.013043     \\
		CRM            & 251            & -0.002059     & 0.098789     & -0.117420    & 0.027698     \\
		AMD            & 251            & -0.001185     & 0.101364     & -0.100103    & 0.037373     \\
		KO             & 251            & 0.000375      & 0.038671     & -0.069626    & 0.012234     \\ \hline
	\end{tabular}
\end{table}

\subsection{Proposed approach}
To demonstrate the comparative effectiveness of our methodology, we utilize Amazon (AMZN) as a case study. The predictive performance of our approach is assessed against other prevalent models, with the findings summarized in Table \ref{tab:model_performance}.

Within the scope of our model performance analysis, it is evident that the Transformer-GAN model excelled, delivering the most accurate predictions. As delineated in Table \ref{tab:model_performance}, this model achieved the lowest Mean Absolute Error (MAE) at 0.0373, the smallest Mean Squared Error (MSE) of 0.0025, and the lowest Normalized Mean Squared Error (NMSE) of 0.2088. These results underscore its superior forecasting capability in comparison to the other models evaluated, including standalone CNN, LSTM, Transformer, and their GAN-enhanced counterparts, with the LSTM-GAN model notably underperforming. Furthermore, the CNN-GAN model underperforms relative to both standalone CNN and LSTM models, indicating that not all mechanisms are compatible with or benefit from integration with GAN structure.

\begin{table}[H]
	\centering
	\caption{Performance Metrics of Predictive Models}
	\label{tab:model_performance}
	\begin{tabular}{lcccc}
		\hline
		\textbf{Model} & \textbf{MAE} & \textbf{MSE} & \textbf{NMSE} \\ \hline
		CNN & 0.0394 & 0.0027 & 0.2237 \\
		LSTM & 0.0414 & 0.0032 & 0.2646 \\
		Transformer & 0.0385 & 0.0026 & 0.2204 \\
		CNN-GAN & 0.0530 & 0.0048 & 0.4003 \\
		LSTM-GAN & 0.1226 & 0.0202 & 1.6920 \\ 
		Transformer-GAN & \textbf{0.0373} & \textbf{0.0025} & \textbf{0.2088} \\ \hline
	\end{tabular}
\end{table}

\subsection{Stock prediction}
We use Transformer-GAN to predice the one-day-ahead stock price, and calculate the one-week-ahead return based on the prediction result. Before the prediction, in order to minimize data modification errors, the values were normalized using $MinMaxScaler$. Then, we calculate the predicted return for all stocks, and the results are shown in Table \ref{tab:weekly_return}. We then select the 10 stocks with the highest predicted returns for inclusion in our portfolio.

\begin{table}[H]
	\centering
	\caption{Weekly Return of Stocks}
	\label{tab:weekly_return}
	\begin{tabular}{l|c}
		\hline
		\textbf{Stock} & \textbf{Predicted Weekly Return} \\
		\hline
		ENPH & 0.2157 \\
		BX & 0.0954 \\
		F & 0.0786 \\
		MSFT & 0.0673 \\
		GOOG & 0.0664 \\
		AAPL & 0.0420 \\
		VZ & 0.0401 \\
		CRM & 0.0351 \\
		AMD & 0.0295 \\
		KO & 0.0249 \\
		PG & 0.0145 \\
		NIO & 0.0133 \\
		NOC & 0.0118 \\
		TSLA & 0.0085 \\
		COST & 0.0055 \\
		ZS & 0.0009 \\
		AMZN & -0.0040 \\
		TSM & -0.0080 \\
		DIS & -0.0085 \\
		INTC & -0.0103 \\
		BA & -0.0133 \\
		META & -0.0325 \\
		PYPL & -0.0514 \\
		XPEV & -0.0627 \\
		NFLX & -0.0690 \\
		\bottomrule
	\end{tabular}
\end{table}

\begin{figure}[H]
	\centering
	\includegraphics[width=\linewidth]{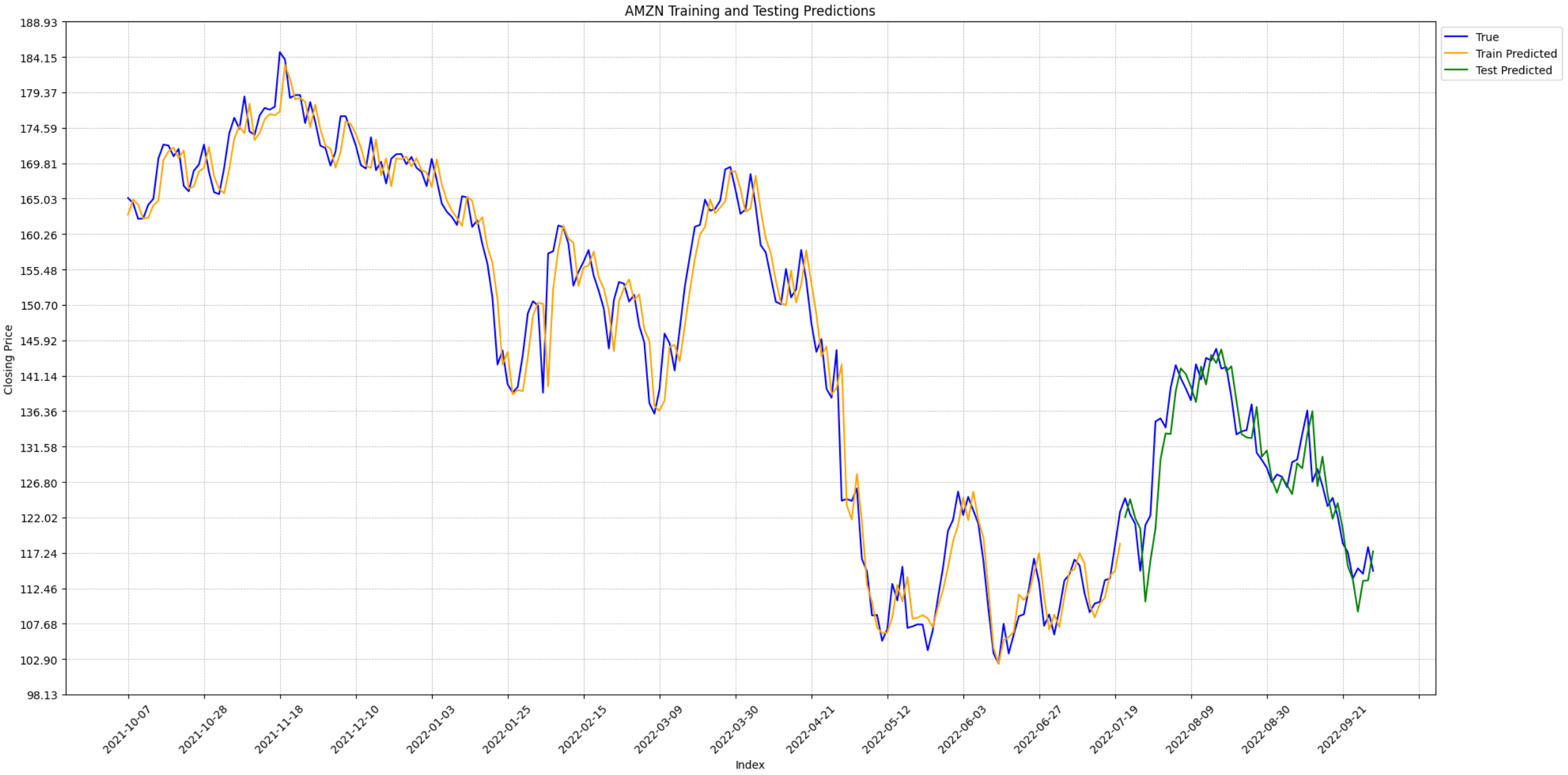}
	\caption{Closing price predictions by the GAN+TRANS model compared with true values over time.}
	\label{fig:gan_trans_results}
\end{figure}

The test results depicted in Fig. \ref{fig:gan_trans_results} demonstrate the Transformer-GAN model's proficiency in forecasting stock prices. The model's predictions, shown in green, closely align with the actual stock prices, which are represented in blue. The overlap of the predicted and true values throughout the timeline signifies the model's robust predictive capability. These findings substantiate the model's potential as an effective tool for financial analysts and investors.

\subsection{Black-Litterman Model Integration}
\subsubsection{Implied Equilibrium Return Vector}
\begin{figure}[htbp]
	\centering
	\includegraphics[width=\linewidth]{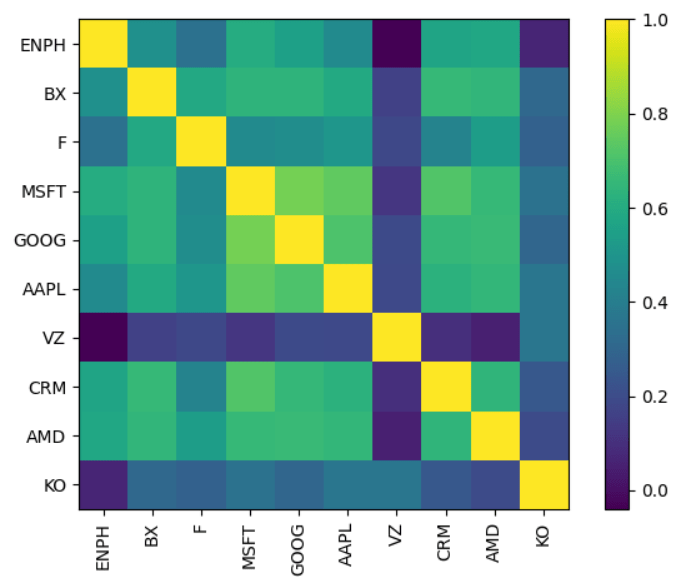}
	\caption{Covariance Matrix Visualization}
	\label{fig:sigma}
\end{figure}

\begin{figure}[htbp]
	\centering
	\includegraphics[width=\linewidth]{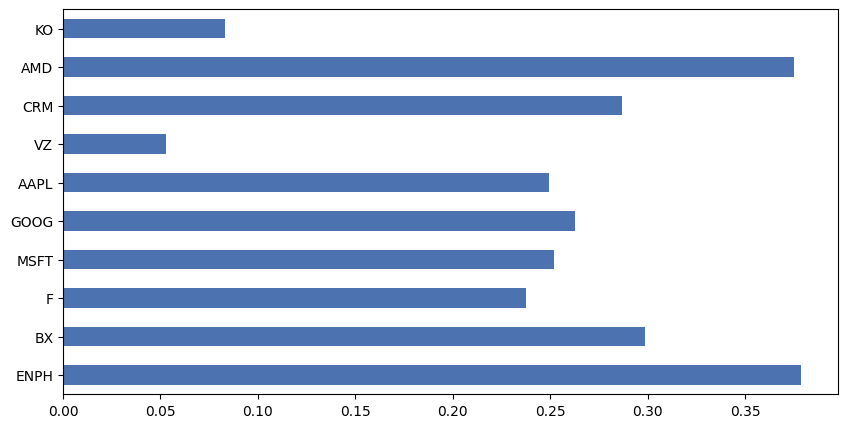}
	\caption{Implied Prior Returns}
	\label{fig:prior}
\end{figure}

Employing market capitalization weights alongside a covariance matrix of returns enabled the calculation of an implied equilibrium return vector, integral to our portfolio optimization strategy. This calculation leverages the proportional market capitals, as detailed in Table \ref{tab:market_cap_weights}, to inform portfolio optimization. Furthermore, the covariance matrix, denoted by $\Sigma$, is illustrated in Fig. \ref{fig:sigma}, providing a visual representation of asset return correlations. Hence, we attain the implied prior returns shown in Fig. \ref{fig:prior}.

\begin{table}[htbp]
	\centering
	\caption{The Market Cap of the 10 Highest Predicted Return}
	\label{tab:market_cap_weights}
	\begin{tabular}{l|c}
		\hline
		\textbf{Stock} & \textbf{Market Cap (USD)} \\
		\hline
		ENPH & 17,686,683,648 \\
		BX & 152,683,495,424 \\
		F & 48,398,815,232 \\
		MSFT & 3,018,393,387,008 \\
		GOOG & 1,688,401,346,560 \\
		AAPL & 2,636,395,315,200 \\
		VZ & 166,110,707,712 \\
		CRM & 296,121,597,952 \\
		AMD & 335,098,675,200 \\
		KO & 256,677,609,472 \\
		\hline
	\end{tabular}
\end{table}

\subsubsection{Constructing Views $(Q)$}
Applying our Transformer-GAN model to predict the one-week-ahead price, the predicted weekly returns are employed as the views for our analysis, as detailed in Table \ref{view}. 
\begin{table}[htbp]
	\centering
	\caption{Views for BL model}
	\label{view}
	\begin{tabular}{l|c}
		\hline
		\textbf{Stock} & \textbf{View} \\
		\hline
		ENPH & 0.2157 \\
		BX & 0.0954 \\
		F & 0.0786 \\
		MSFT & 0.0673 \\
		GOOG & 0.0664 \\
		AAPL & 0.0420 \\
		VZ & 0.0401 \\
		CRM & 0.0351 \\
		AMD & 0.0295 \\
		KO & 0.0249 \\
		\hline
	\end{tabular}
\end{table}

\subsection{Portfolio Optimization}
In order to validate the efficiency of our proposed method in portfolio construction, we applied various optimization strategies. This includes the traditional mean-variance (MV) approach, the equal-weighted strategy, and the Black-Litterman (BL) model with analyst-derived views. The performance of these strategies is measured against our method to highlight its superior adaptability and effectiveness. 

Table \ref{tab:portfolio_optimization} and Fig. \ref{fig:adjusted} below displays the Prior, Posterior, and Views for a selection of stocks. This data showcases how the BL model adjusts the original market expectations to align more closely with the expectation of our model, allowing for a more informed portfolio construction that potentially leads to enhanced performance.

\begin{table}[htbp]
	\centering
	\caption{Adjusted returns with views}
	\label{tab:portfolio_optimization}
	\begin{tabular}{lccc}
		\hline
		\textbf{Stock} & \textbf{Prior} & \textbf{Posterior} & \textbf{Views} \\ \hline
		ENPH & 0.378782 & 0.161047 & 0.215665 \\
		BX   & 0.298731 & 0.101838 & 0.095407 \\
		F    & 0.237640 & 0.082635 & 0.078559 \\
		MSFT & 0.252061 & 0.101625 & 0.067298 \\
		GOOG & 0.262989 & 0.102187 & 0.066403 \\
		AAPL & 0.249424 & 0.094659 & 0.041993 \\
		VZ   & 0.053028 & 0.034654 & 0.040120 \\
		CRM  & 0.286891 & 0.090919 & 0.035144 \\
		AMD  & 0.374907 & 0.111667 & 0.029542 \\
		KO   & 0.083053 & 0.035655 & 0.024880 \\ \hline
	\end{tabular}
\end{table}

\begin{figure}[htbp]
	\centering
	\includegraphics[width=\linewidth]{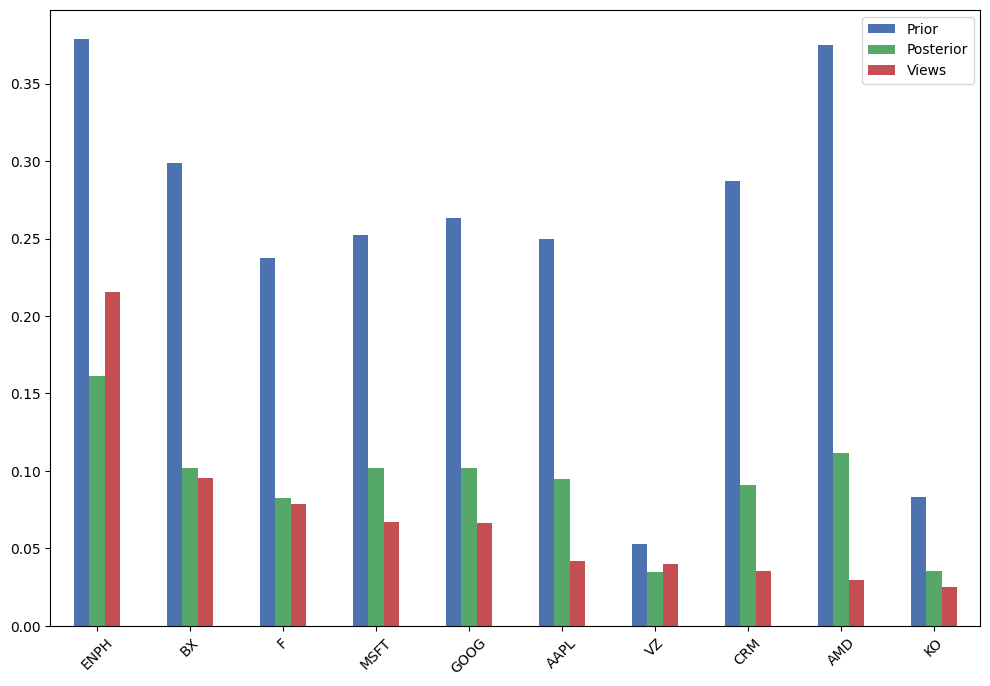}
	\caption{Adjusted returns with views}
	\label{fig:adjusted}
\end{figure}

\begin{figure}[htbp]
	\centering
	\includegraphics[width=\linewidth]{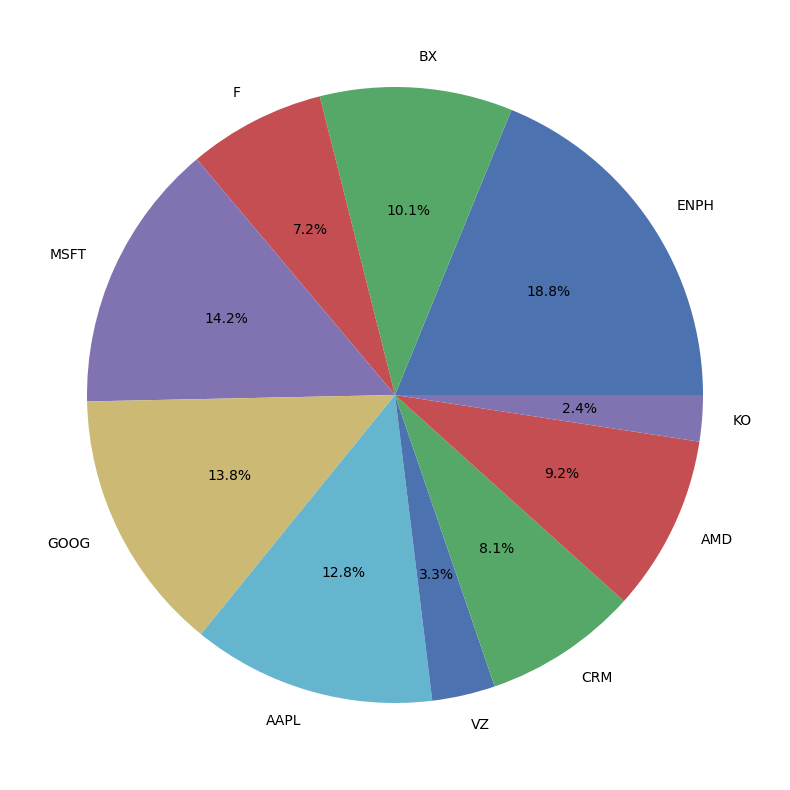}
	\caption{Weights (Our method)}
	\label{fig:wei_bl}
\end{figure}

Fig. \ref{fig:wei_bl} illustrates the weights derived from our method, while Table \ref{tab:portfolio_weights} displays the weights calculated using our method, mean-variance and equal-weighted strategies, respectively.

\begin{table}[htbp]
	\centering
	\caption{Portfolio Weights by Different Optimization Models}
	\label{tab:portfolio_weights}
	\begin{tabular}{@{}lccc@{}}
		\hline
		\textbf{Stock} & \textbf{Our Method} & \textbf{Mean-Variance} & \textbf{Equal Weights} \\
		\hline
		AAPL & 12.8\% & 12.0\% & 10.0\% \\
		MSFT & 14.2\% & 11.8\% & 10.0\% \\
		GOOG & 13.8\% & 12.1\% & 10.0\% \\
		ENPH & 18.8\% & 12.4\% & 10.0\% \\
		KO & 2.4\% & 3.5\% & 10.0\% \\
		AMD & 9.2\% & 14.5\% & 10.0\% \\
		CRM & 8.1\% & 11.7\% & 10.0\% \\
		VZ & 3.3\% & 2.0\% & 10.0\% \\
		BX & 10.1\% & 11.4\% & 10.0\% \\
		F & 7.2\% & 8.4\% & 10.0\% \\
		\hline
	\end{tabular}
\end{table}

\begin{figure*}[htbp]
	\centering
	\includegraphics[width=\linewidth]{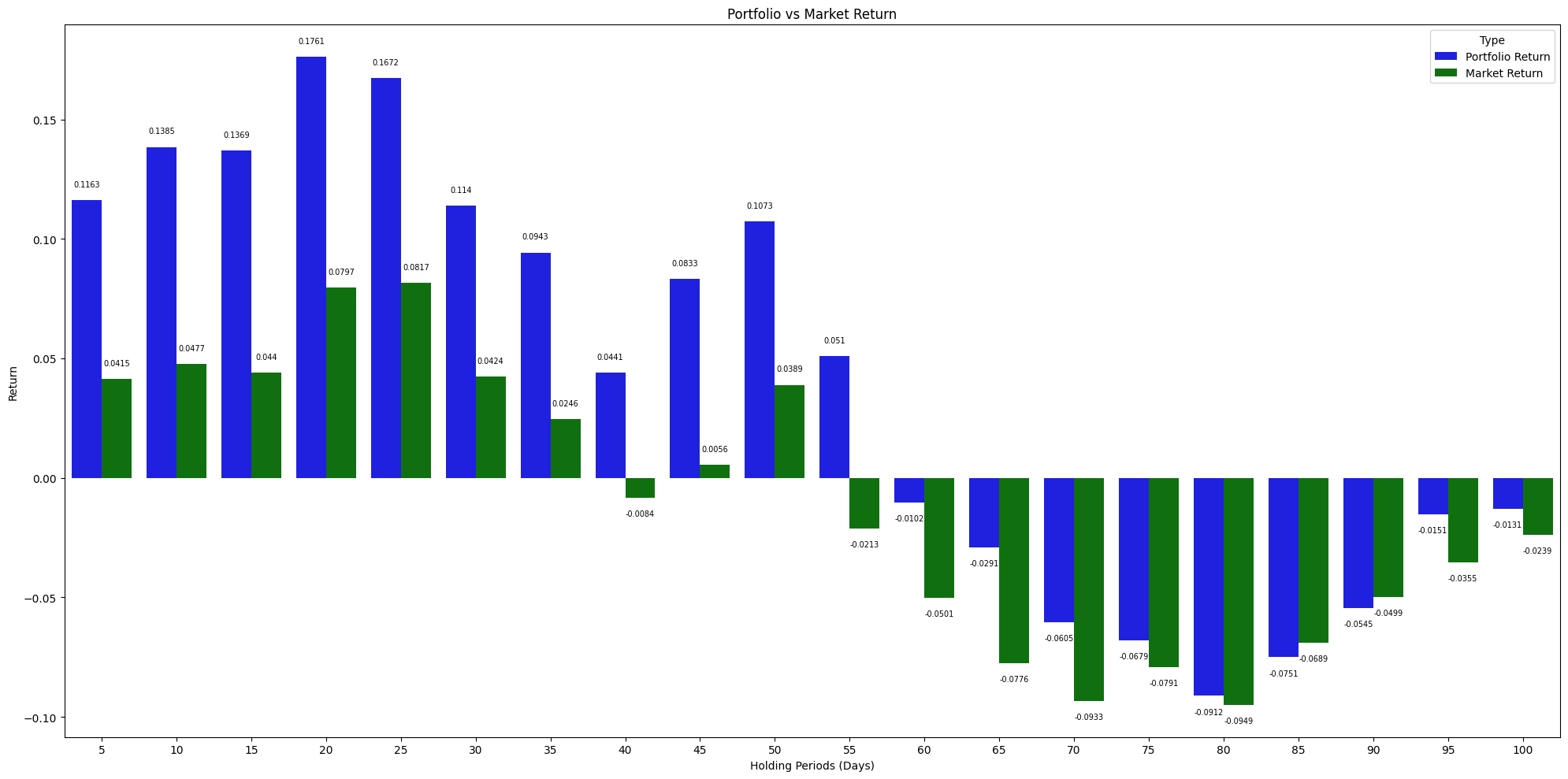}
	\caption{Returns comparison for different periods (Our method)}
	\label{bl_novel}
\end{figure*}

\begin{figure}[H]
	\centering
	\includegraphics[width=\linewidth]{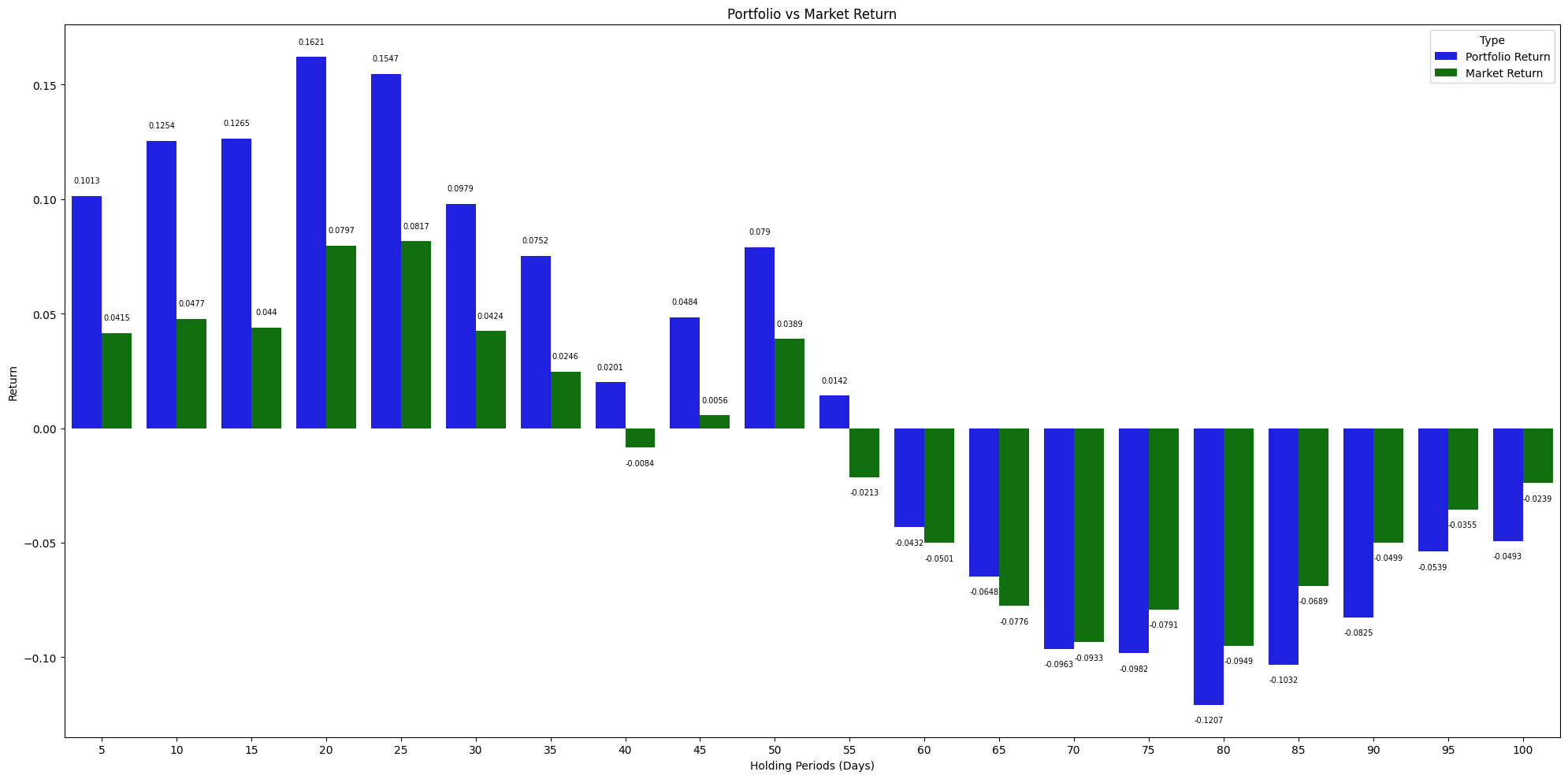}
	\caption{Returns comparison for different periods (MV)}
	\label{mv}
\end{figure}

\begin{figure}[H]
	\centering
	\includegraphics[width=\linewidth]{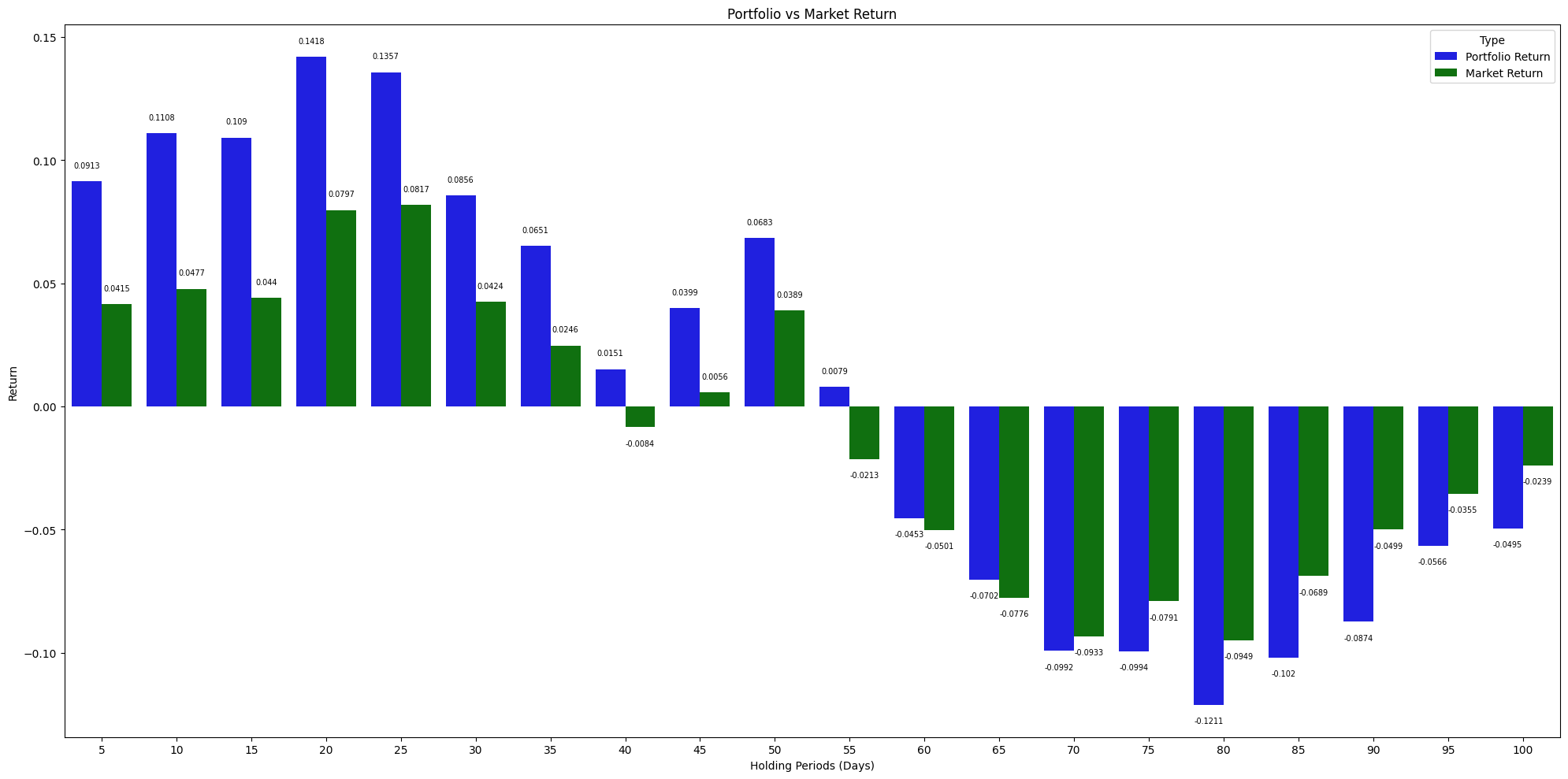}
	\caption{Returns comparison for different periods (Equal weights)}
	\label{ew}
\end{figure}

\begin{figure}[H]
	\centering
	\includegraphics[width=\linewidth]{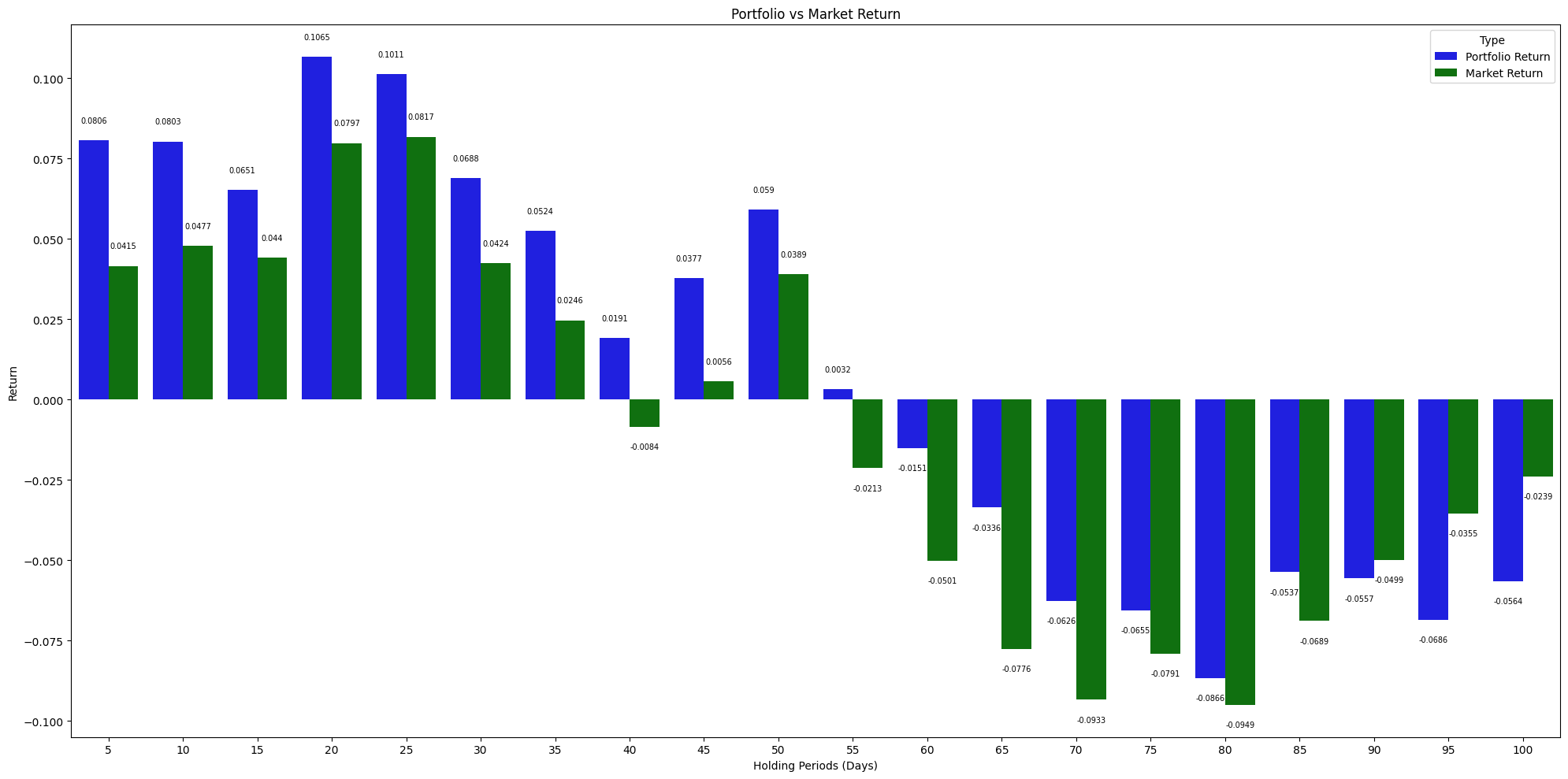}
	\caption{Returns comparison for different periods (BL:view from analyst 1)}
	\label{bl_1}
\end{figure}

\begin{figure}[H]
	\centering
	\includegraphics[width=\linewidth]{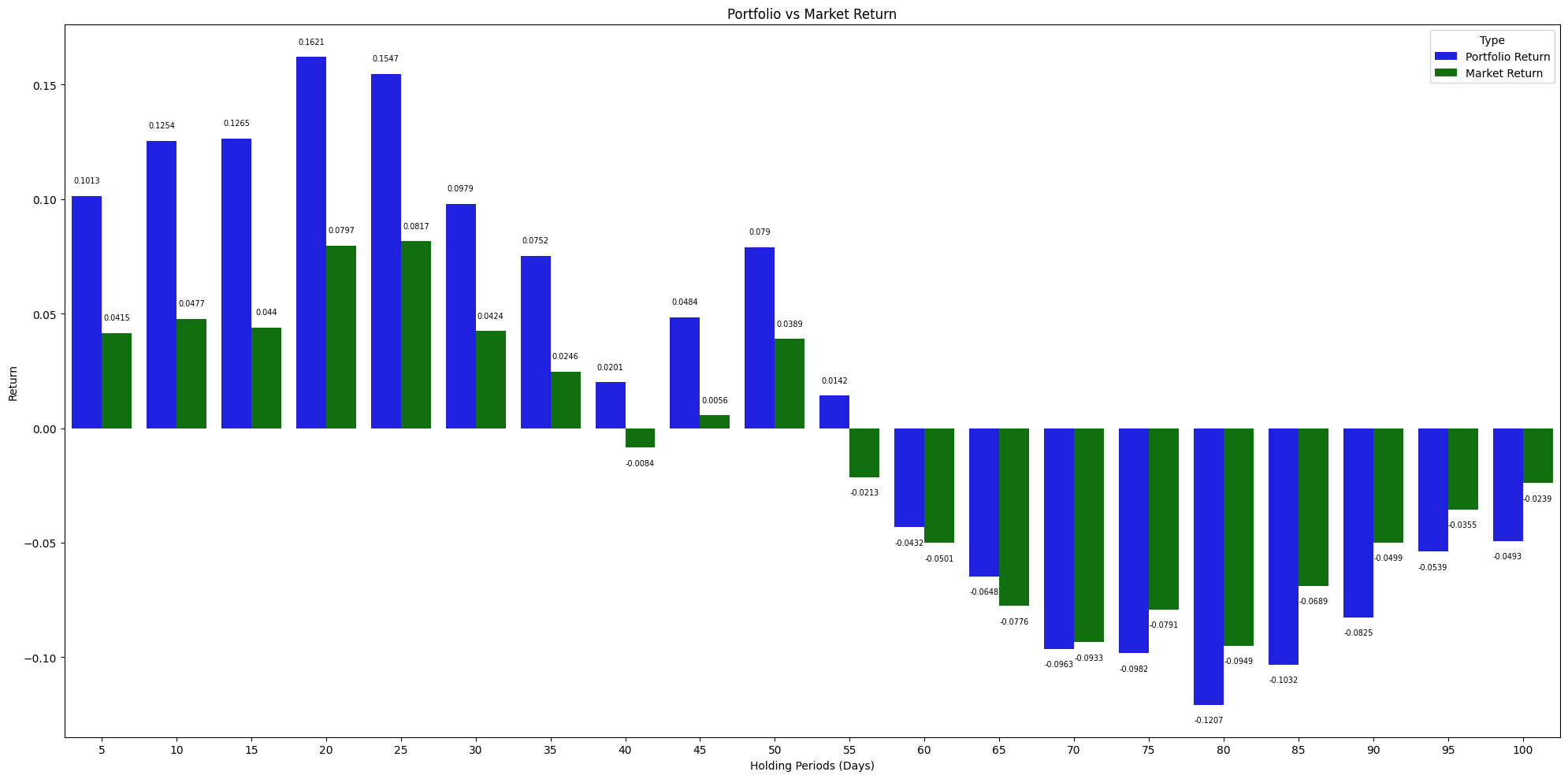}
	\caption{Returns comparison for different periods (BL:view from analyst 2)}
	\label{bl_2}
\end{figure}

Fig. \ref{bl_novel}, \ref{mv}, and \ref{ew} display the returns for various holding periods under different optimization strategies. Fig. \ref{bl_novel} illustrates the returns using our method, consistently outperforming the benchmarks, particularly for longer holding periods. This is in contrast to Fig. \ref{mv} and \ref{ew}, which represent the MV and equal-weighted strategies, respectively. For the Black-Litterman strategy, Fig. \ref{bl_1} and \ref{bl_2} present outcomes generated based on views from analysts. These figures show a competitive start but generally underperform compared to our method, especially as the holding period increases. In the later stages, the observed minimal discrepancy between the market return and the return generated by our methodology can be attributed to typical market fluctuation which is a common situation.

\begin{figure*}[htbp]
	\centering
	\includegraphics[width=\linewidth]{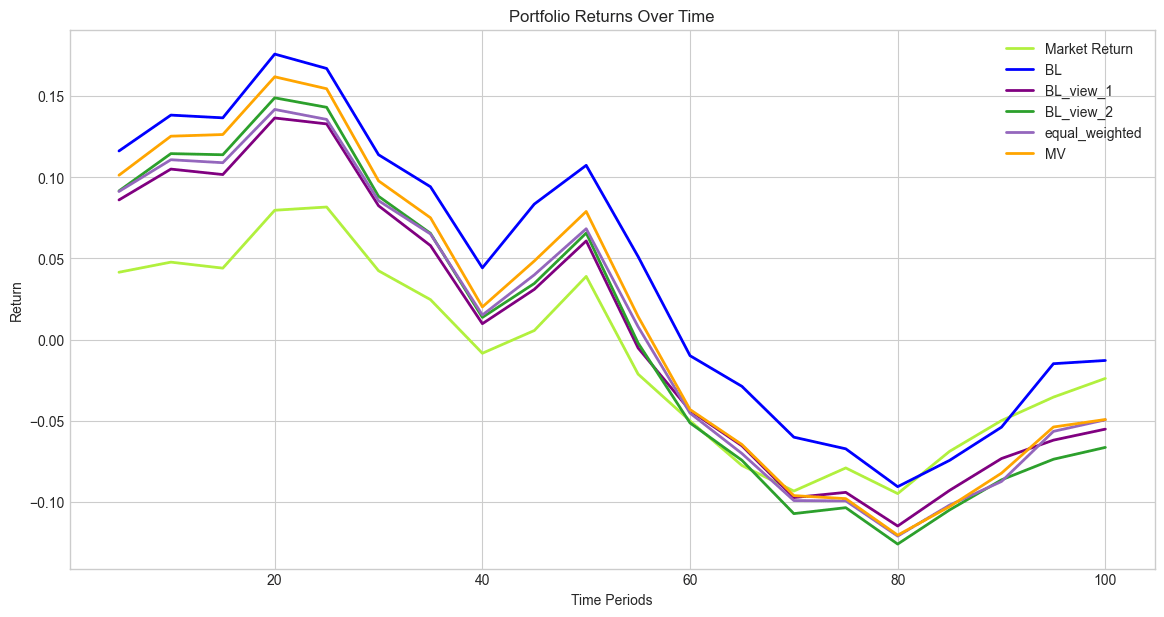}
	\caption{Comparative Analysis of Portfolio Returns Across Different Investment Strategies}
	\label{com}
\end{figure*}

The graphical representation in Fig. \ref{com} provides a comprehensive overview of portfolio returns over various time periods, contrasting the performance of different investment strategies against actual market returns. Our method, indicated by the \textbf{Blue} line which is marked as \textbf{`BL'}, exhibits robust returns that surpass the performance of all strategies for the majority of the observed periods. Notably, in the later stages, the market return, illustrated by the orange line, outperforms all the strategies, including ours. This observation is consistent with the typical market behavior, where long-term fluctuations may favor the actual market returns over modeled strategies \cite{svedsater2009momentum}\cite{malin2013long}\cite{campbell2005term}. Such trends underscore the challenges in outperforming the market consistently, especially in later periods where market dynamics may change rapidly. Despite this, the superior performance of our model in the earlier phases demonstrates its potential to provide investors with significant insights for strategic decision-making in asset management.

\begin{figure*}[htbp]
	\centering
	\begin{subfigure}{.5\textwidth}
		\centering
		\includegraphics[width=.8\linewidth]{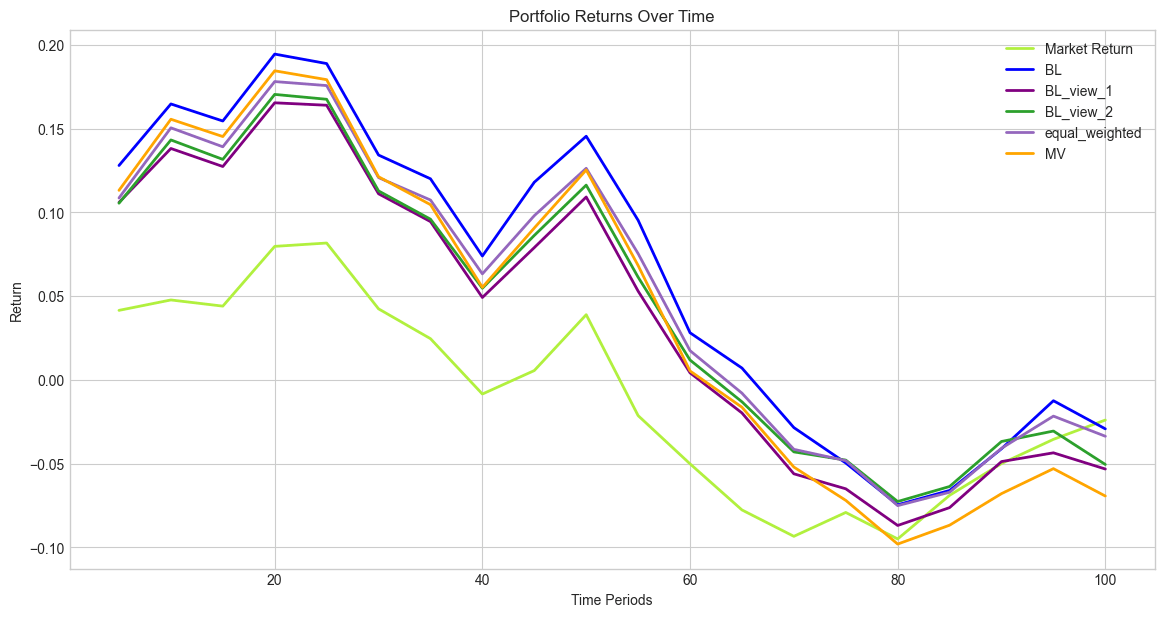}
		\caption{Random Entry Analysis I}
		\label{fig:strategy_perf_1}
	\end{subfigure}%
	\begin{subfigure}{.5\textwidth}
		\centering
		\includegraphics[width=.8\linewidth]{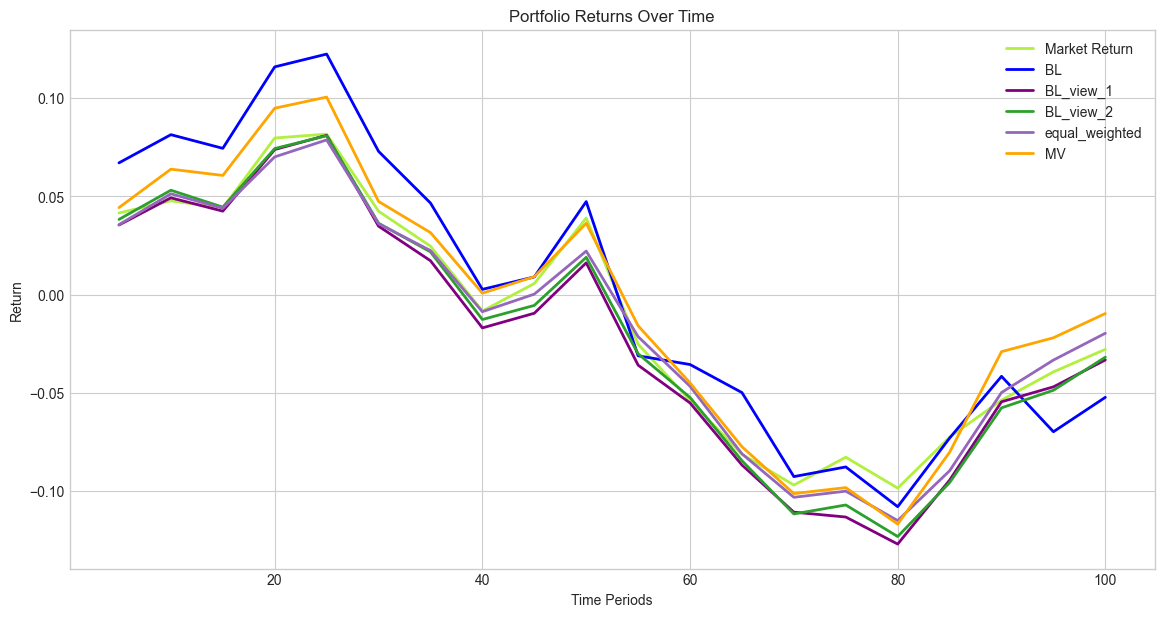}
		\caption{Random Entry Analysis II}
		\label{fig:strategy_perf_2}
	\end{subfigure}
	\begin{subfigure}{.5\textwidth}
		\centering
		\includegraphics[width=.8\linewidth]{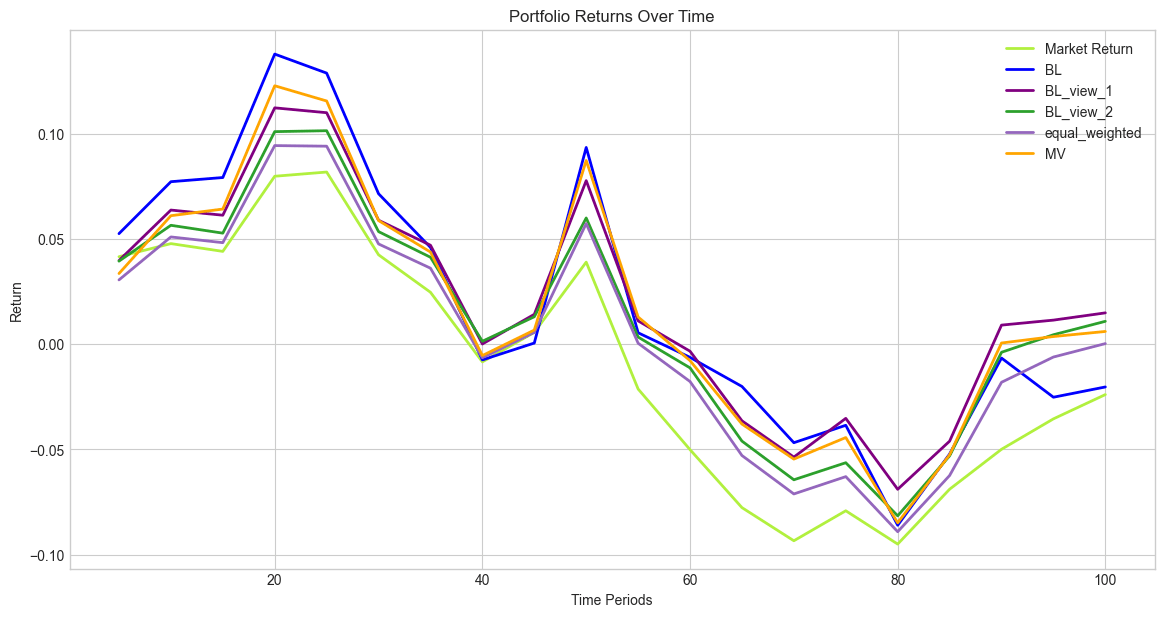}
		\caption{Random Entry Analysis III}
		\label{fig:strategy_perf_3}
	\end{subfigure}%
	\begin{subfigure}{.5\textwidth}
		\centering
		\includegraphics[width=.8\linewidth]{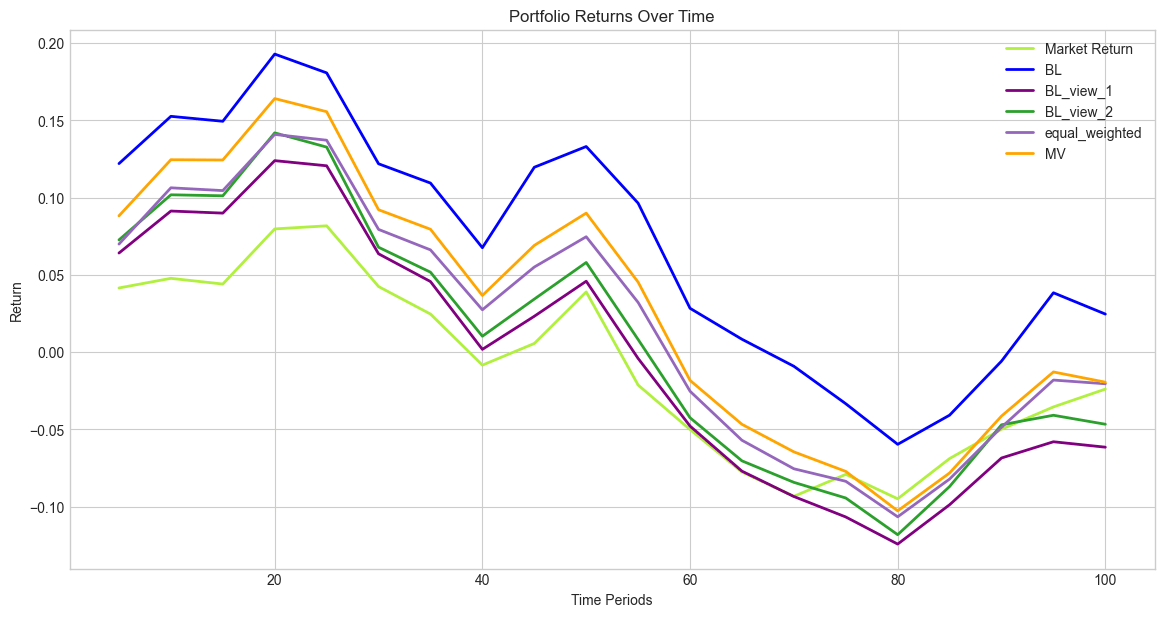}
		\caption{Random Entry Analysis IV}
		\label{fig:strategy_perf_4}
	\end{subfigure}
	\caption{Stochastic Entry Points and Portfolio Performance}
	\label{fig:stochastic_performance}
\end{figure*}

Fig. \ref{fig:stochastic_performance} offers a detailed visualization of portfolio returns corresponding to four distinct scenarios, each scenario characterized by a set of randomly selected entry points. The ensemble of these subfigures demonstrates the robust performance of the proposed investment strategy across a spectrum of stochastic market conditions. Despite the inherent unpredictability of entry points, the strategy exhibits a persistent outperformance, highlighting its potential for strategic asset management.

\section{Conclusion}
Recent advancements in AI, particularly the popularity of attention mechanisms and the use of LSTM and CNN models for financial forecasting, have significantly improved the predictive accuracy of stock price movements. Building upon these advancements, our study introduces a novel approach by integrating Transformer models—known for their exceptional ability to capture long-range dependencies within data—with Generative Adversarial Networks (GANs). This combination allows our model to not only predict stock prices with high accuracy but also validate these predictions against actual historical data.

The implications of our findings extend beyond the academic sphere, offering tangible benefits to portfolio managers and financial analysts. By incorporating our Transformer-GAN model, practitioners can achieve more accurate predictions of market movements and asset prices, facilitating better decision-making and risk management. Theoretically, this research enriches the dialogue on the utility of machine learning in finance, particularly highlighting the benefits of combining different types of neural networks to address specific financial tasks.

In conclusion, our transformative approach leverages a Transformer-GAN model to generate predictive views for the Black-Litterman model, achieving novel results in portfolio optimization. This integration not only enhances predictive accuracy but also innovates on traditional financial models by incorporating cutting-edge AI technologies. Our findings represent a significant step forward in the application of machine learning in finance, offering a sophisticated method for asset allocation that aligns with modern investment strategies and market expectations. 

Despite its promising outcomes, our approach faces challenges, including sensitivity to market volatility, data overfitting risks, scalability concerns due to large data dependencies, and GANs' training instability leading to convergence issues. Innovative solutions are needed to enhance predictive accuracy while maintaining efficiency. Future work may explore hybrid deep learning models to address these limitations, extend the model's application across various financial instruments and markets, and improve computational efficiency for broader industry adoption.

\section*{Acknowledgment}

This work is supported by the Guangdong Basic and Applied Basic Research Foundation (No. 2023B1515130002).

\bibliography{ref}
\bibliographystyle{IEEEtran}
\end{document}